\documentclass[12pt]{article}
\usepackage{epsf}
\usepackage[utf8]{inputenc}
\usepackage[english]{babel}

\begin{document}

\title{
{\bf Comments to the paper by L. Durand and P. Ha on Coulomb-nuclear interference. }}
\author{ Vladimir A. Petrov \thanks{E-mail: Vladimir.Petrov@ihep.ru}\\
{\small {\it A.A. Logunov Institute for High Energy Physics}},\\ {\small {\it NRC ``Kurchatov Institute'', 142281 Protvino, Russia}}}
\date{}
\maketitle

\vskip-1.0cm

\begin{abstract}
This is a brief analysis of the basic formula from  Ref.\cite{Du1}. Some inconsistency has been identified.
\end{abstract}

In the recent paper \cite{Du1}, there was exhibited quite a thorough analysis of the data on proton-proton elastic scattering data with an emphasis on the estimate of the nuclear phase parameter $ \rho (s, q^{2})= Re f_{N}(s,q^{2})/Im f_{N}(s,q^{2}) $ with $ f_{N}(s,q^{2}) $ is the strong interaction scattering amplitude. Besides the "practical" importance of the parameter $ \rho  $ (e.g. for the total cross-sections retrieval) the conceptual importance of the $ q^{2} $-dependence of the strong interaction phase has been explained in Ref.\cite{Ku1} (see also \cite{Pe1}). 

In the framework of the eikonal scheme with an eikonal additive in electromagnetic and strong interactions the authors of Ref. \cite{Du1} brought the total amplitude of proton-proton scattering $ f(s,q^{2}) $ to the form (see Eq.(31) in \cite{Du1}) 
\begin{equation}
f(s,q^{2})= f_{C}^{B} + e^{i\Phi_{tot}(s,q^{2})}f_{N}(s,q^{2})
\end{equation}
where $ f_{C}^{B} \sim \alpha F_{Q}^{2}(q^{2})/q^{2} $ is the Coulomb Born amplitude ($F_{Q}^{2}(q^{2})$ is the proton charge form factor ), 
\[f_{N}(s,q^{2})=\mid f_{N}(s,q^{2})\mid e^{i\Phi_{N}(s,q^{2})}\]
is the pure strong interaction ("nuclear") scattering amplitude while the phase  $ \Phi_{tot}(s,q^{2}) $ reflects the joint contributions of the strong and electromagnetic interactions.

In spite of quite different arguments when deriving the phase $ \Phi_{tot}(s,q^{2}) $ in 
Ref.\cite{Du1} the  form of the full scattering amplitude  (1) is functionally equivalent to the form first proposed by H. Bethe in Ref.\cite{Be}

\begin{equation}
f(s,q^{2})= f_{C}^{B} + e^{i\alpha \Phi_{Bethe}(s,q^{2})}f_{N}(s,q^{2})
\end{equation}
 because the phase
$ \Phi_{tot}(s,q^{2})\sim \alpha $.

 In fact, when we switch off electromagnetic interaction we should get 
\[\lim_{\alpha \rightarrow 0} f(s,q^{2})= f_{N}(s,q^{2}). \]

It was noticed long ago ( see the first item in \cite{Ku1}) that the Bethe form (1) ((2)) of the full amplitude necessarily requires that the phase of the nuclear amplitude $ \Phi_{N}(s,q^{2}) $
does not depend on $ q^{2} $. Actually, it was already evident much earlier from the expression for the interference phase obtained in \cite{Ca}.

 However,  it was \textit{assumed} in Ref.\cite{Du1}, for reasons of a better description of the data, that 
 \[\rho (s, q^{2})= Ref_{N}(s,q^{2})/Im f_{N}(s,q^{2})\approx \rho (s)\frac{1-q^{2}/q^{2}_{R}}{1-q^{2}/q^{2}_{I}} .\]
 
 This formula, in its turn, implies that 
 \begin{equation}
 \Phi_{N}(s, q^{2}) = \arctan [ \frac{1}{\rho (s)}\frac{1-q^{2}/q^{2}_{I}}{1-q^{2}/q^{2}_{R}} ].
 \end{equation}
However, such an evident $ q^{2} $-dependence of the nuclear phase is not compatible, as was shown in Refs. \cite{Ku1}, \cite{Cu2}, \cite{Pe2}, with a Bethe-like form (1) of the full amplitude $  f(s,q^{2}) $.

This statement is not quite evident therefore we believe it is appropriate to give
its simple proof.

To this end let us compare the values of the moduli squared of the full amplitude as in Eq.(1) and that which follows from the additive eikonal scheme. Eq.(1) gives
\begin{equation}
\mid f(s,q^{2}) \mid^{2}= \mid f_{N}(s,q^{2}) \mid^{2} + \mid f_{C}^{B} \mid^{2} +
\end{equation}
\[ + 2f_{C}^{B}\mid f_{N}(s,q^{2}) \mid \cos(\Phi_{N}(s,q^{2})+\Phi_{tot}(s,q^{2}))\]
 Let us now take Eq.(4) up to the first order in $ \alpha $ (take note that $ f_{C}^{B}\sim \alpha $):
 \begin{equation}
 \mid f(s,q^{2}) \mid^{2}= \mid f_{N}(s,q^{2}) \mid^{2} +2f_{C}^{B}\mid f_{N}(s,q^{2}) \mid \cos(\Phi_{N}(s,q^{2})) + \mathcal{O} (\alpha^{2})
 \end{equation}
It is not difficult to show that in the additive eikonal scheme (which was claimed to be the case in \cite{Du1}) the following general expression for $\mid f(s,q^{2}) \mid^{2}$ holds:
\begin{equation}
 \mid f(s,q^{2}) \mid^{2}= \mid f_{N}(s,q^{2}) \mid^{2} +2f_{C}^{B}\mid f_{N}(s,q^{2}) \mid \cos(\Phi_{N}(s,q^{2})) -
\end{equation}
\[- \frac{\alpha}{\pi} \int \frac{d^{2}k}{k^{2}}F_{Q}^{2}(k^{2})Im[f_{N}(s,q^{2})f_{N}^{*}(s,(q-k)^{2})] + \mathcal{O} (\alpha^{2}).\]

Hence it follows from Eqs. (5) and (6) that
\begin{equation}
\int \frac{d^{2}k}{k^{2}}F_{Q}^{2}(k^{2})Im[f_{N}(s,q^{2})f_{N}^{*}(s,(q-k)^{2})] = 0 
\end{equation}
and then
\begin{equation}
\ \int \frac{d^{2}k }{k^{2}}F_{Q}^{2}(k^{2}) Im[f_{N}(s,(q-k)^{2}) [\rho (q^{2})-\rho ((k-q)^{2}])=0.
\end{equation}

As Eq.(8)holds at arbitrary $ q^{2} $ and, evidently, 
\[\frac{F_{Q}^{2}(k^{2})Im[f_{N}(s,(q-k)^{2})}{k^{2}} \neq \delta (\textbf{k}),\]
this leaves no other choice as independence of  $ \rho $ on the momentum transfer 
which does not comply with Eq.(3).

Comparison of $ \mathcal{O} (\alpha^{2}) $ terms yields the expression for $ \Phi_{tot}(s,q^{2}) $ in terms of $ f_{N}(s,q^{2}) $ and $ f_{C}^{B} $. It should be noted that it was shown long ago in the first of references \cite{Ku1} that reality of the resulting expression for $ \Phi_{tot} $  leads again to independence of $ \Phi_{N} $ from $ q^{2} $.

Comparison of higher powers in $ \alpha $ would lead to an infinite number of constraints on $ f_N $ whose feasibility is hardly possible. 

\section*{Conclusions}

 Our  conclusion is as follows:
 
  Eq.(3)( Eq.(37) in the paper by L. Durand and P.Ha \cite{Du1}) contradicts the basic premises  (eikonal additivity \textit{and} a Bethe-like parametrization of the full amplitude) of the model suggested in \cite{{Du1}}. 
 
 We believe that in view of the popularity of the parametrization of the Coulomb-nuclear interference in a form that goes back to Bethe (see eq. (1) or (2)), it is necessary to clearly understand when using it that in this case the additivity of the eikonal with respect to strong and electromagnetic interactions and non-trivial dependence of the nuclear phase on momentum transfer ($ \rho(q^{2}) \neq \textit{const}) $ are incompatible.
 Any model which ignores this circumstance is self-contradictory although at first glance it may look quite decent phenomenologically (good chi-square, etc.)
 This is exactly the case of the Durand-Ha model \cite{Du1}.
 
\section*{Acknowledgement}
I am grateful to Prof. L. Durand for stimulating correspondence concerning this issue.

\end{document}